\documentclass{ws-ijmpd}

\newcommand{\rr}{\mbox{\boldmath $r$}}

\newcommand{\rk}{\mbox{\boldmath $k$}}

\newcommand{\rkn}{\mbox{$k$}}
\newcommand{\rp}{\mbox{\boldmath $p$}}

\begin{document}

\markboth{V.P. Gon\c{c}alves and M. V. T. Machado}
{Heavy Quarks and Meson Production in Ultraperipheral Heavy Ion Collisions}

\catchline{}{}{}{}{}
%
%%%%%%%%%%%%%%%%%%%%%%%%%%%%%%%%%%%%%%%%%%%%%%%%%%%%%%%%%%%%%%%%%%%%

\title{Heavy Quarks and Meson Production in Ultraperipheral Heavy Ion Collisions }

\author{V. P. Gon\c{c}alves}

\address{Instituto de F\'{\i}sica e Matem\'atica, Universidade Federal de
Pelotas\\
Caixa Postal 354, CEP 96010-090, Pelotas, RS, Brazil \\
barros@ufpel.edu.br}

\author{M. V. T. Machado}

\address{Universidade Estadual do Rio Grande do Sul (UERGS). Unidade de Bento Gon\c{c}alves \\ Rua Benjamin Constant, 229. Bento Gon\c{c}alves. CEP 95700-000, Brazil
\\
magno-machado@uergs.edu.br}

\maketitle

\begin{history}
\received{Day Month Year}
\revised{Day Month Year}
\comby{Managing Editor}
\end{history}

\begin{abstract}
We report on  our recent  investigations in photonuclear production of   heavy quarks and vector meson in
ultraperipheral heavy ion collisions. In particular, our theoretical predictions are compared with the recent experimental measurements on coherent $\rho$ (STAR) and $J/\Psi$ (PHENIX) photoproduction at RHIC and estimates for LHC are given.

\end{abstract}


%\section{General Appearance}
\section{Introduction}

In the last years, there has been a lot of interest in the
description of electron-nucleus collisions at high energies.
The results of current  analysis show that future  electron-nucleus
colliders at HERA and RHIC, probably could
determine whether parton distributions saturate and  constraint
the behavior of the nuclear gluon distribution in the full
kinematical range \cite{VPG,VPG2}. However, until these colliders become reality
we need to consider alternative searches  in the current and/or
scheduled  accelerators which allow us to constraint the QCD
dynamics. In this contribution, we summarize our investigations in Refs. \cite{EPJCPER3,per4} about the  possibility of using
ultraperipheral heavy ion collisions (UPC's) as a photonuclear collider
and study the heavy quark and vector meson production  assuming distinct approaches
for the QCD evolution.

%\section{Photonuclear processes at heavy ion collisions}

In heavy ion  collisions the large number of photons coming from
one of the colliding nuclei  will  allow to study photoproduction,
with energies $W_{\gamma A}$ reaching to  950 GeV for the LHC. The
photonuclear cross sections are given by the convolution between
the photon flux from one of the nuclei and the cross section for
the scattering photon-nuclei.  The expression
for the production of a given final state $Y$ ($Y = Q \overline{Q}, \,V$) ultraperipheral heavy ion
collisions is then given by,
\begin{eqnarray}
\sigma_{AA \rightarrow A Y X}\,\left(\sqrt{S_{\mathrm{NN}}}\right) = \int \limits_{\omega_{min}}^{\infty} d\omega \, \frac{dN\,(\omega)}{d\omega}\,\, \sigma_{\gamma A \rightarrow YX} \left(W_{\gamma A}^2=2\,\omega\sqrt{S_{\mathrm{NN}}}\right)\,
\label{sigAA}
\end{eqnarray}
where $\omega_{min}=M_{Y}^2/4\gamma_L m_p$, 
$\sqrt{S_{\mathrm{NN}}}$ is  the c.m.s energy of the
nucleus-nucleus system and  $\gamma_L$ is the Lorentz boost  of a single beam.  The Lorentz factor for LHC is
$\gamma_L=2930$, giving the maximum c.m.s. $\gamma A$ energy
$W_{\gamma A} \leq 950$ GeV.  The requirement that  photoproduction
is not accompanied by hadronic interaction (ultraperipheral
collision) can be done by restricting the impact parameter $b$  to
be larger than twice the nuclear radius, $R_A=1.2 \,A^{1/3}$ fm. An analytic approximation for $AA$ collisions can be obtained
using as integration limit $b>2\,R_A$, producing
\begin{eqnarray}
\frac{dN\,(\omega)}{d\omega}= \frac{2\,Z^2\alpha_{em}}{\pi\,\omega}\, \left[\bar{\eta}\,K_0\,(\bar{\eta})\, K_1\,(\bar{\eta})+ \frac{\bar{\eta}^2}{2}\,\left(K_1^2\,(\bar{\eta})-  K_0^2\,(\bar{\eta}) \right) \right] \,
\label{fluxint}
\end{eqnarray}
where $\bar{\eta}=2\omega\,R_A/\gamma_L$ and $K_{0,1}(x)$ are the modified Bessel functions. Therefore, the main ingredient for computing the production of the specific final state $Y$ at UPC's is the information about its cross section in photon-nuclei interactions, $\sigma_{\gamma A\rightarrow YX}$, as defined in Eq. \ref{sigAA}. In what follows, we perform phenomenology on heavy quarks and vector mesons production using the available high energy approaches.

\section{Photonuclear  production of heavy
quarks} 

For our further analysis on photonuclear  production of heavy
quarks we will consider distinct high energy approaches, which are contrasted against the collinear approach 
\cite{EPJCPER3}: (i) semihard formalism, (ii) saturation model (within the color dipole formalism) and (iii) Color Glass Condensate (CGC) approach. The main goal is a comparison among those approaches and show whether the production at UPC's allows disentangle the underlying QCD dynamics at high energies.

(i) \underline{\bf Semihard formalism}: In the $k_{\perp}$-factorization (semihard) approach, the
relevant QCD diagrams are considered with the virtualities and
polarizations of the initial partons, carrying information on
their transverse momenta. The scattering processes are described
through the convolution of off-shell matrix elements with the
unintegrated parton distribution, ${\cal F}(x,\rk_{\perp})$. Considering
only the direct component of the photon, the cross section reads as (See, e.g. Ref. \cite{SMALLX}),
\begin{eqnarray}
\sigma_{\gamma A\rightarrow Q\bar{Q}X}^{\mathrm{semihard}} &= &
  \frac{\alpha_{em}\,e_Q^2}{\pi}\, \int\, dz\,\,d^2 \rp_{1\perp} \, d^2\rk_{\perp} \, \frac{\alpha_s\,{\cal F}_{\mathrm{nuc}}(x,\rk_{\perp}^2)}{\rk_{\perp}^2}
 \left\{ A(z)\left( \frac{\rp_{1\perp}}{D_1} + \frac{(\rk_{\perp}-\rp_{1\perp})}{D_2} \right)^2 \right. \nonumber \\
& & + \left.  m_Q^2 \,\left(\frac{1}{D_1} + \frac{1}{D_2}  \right)^2  \right\}
 \label{sigmakt}
\end{eqnarray}
where $D_1 \equiv \rp_{1\perp}^2 + m_Q^2$ and $D_2 \equiv (\rk_{\perp}-\rp _{1\perp})^2 + m_Q^2$ and $A(z)=[z^2+ (1-z)^2] $. The transverse momenta of the heavy quark (antiquark) are denoted by $\rp_{1\perp}$ and $\rp_{2\perp}= (\rk_{\perp}-\rp _{1\perp})$, respectively. The heavy quark longitudinal momentum fraction is labeled by $z$. For the scale $\mu$ in the strong coupling constant we  use the prescription  $\mu^2=\rk_{\perp}^2 + m_{Q}^2$. We use also the simple ansatz for the unintegrated gluon distributions, ${\cal F}_{\mathrm{nuc}} =  \frac{\partial\, xG_{A}(x,\,\rk_{\perp}^2)}{\partial \ln \rk_{\perp}^2}$
where $xG_{A}(x,Q^2)$ is the nuclear gluon distribution (see Ref.\cite{EPJCPER3} for details  in the  numerical calculations).

(ii) \underline{\bf Saturation model}: Based on the color dipole formalism, it is an extension of the $ep$
saturation model through  Glauber-Gribov formalism. In this model
the cross section for the  heavy quark photoproduction on nuclei
targets is given by \cite{armesto,nik91},
\begin{eqnarray}
\sigma_{\gamma A\rightarrow Q\bar{Q}X}^{\mathrm{dipole}} = \int_0^1
dz\, \int d^2\rr \, |\Psi_{T} (z,\,\rr,\,Q^2=0)|^2 \, \sigma_{dip}^{\mathrm{A}}
(\tilde{x},\,\rr^2,A)\,,
\label{sigmaphot}
\end{eqnarray}
where the transverse wave function is known \cite{nik91,predazzi}.  The nuclear dipole cross section is given by \cite{armesto,nik91},
\begin{eqnarray}
\sigma_{dip}^{\mathrm{A}} (\tilde{x}, \,\rr^2, A)  = \int d^2b \,\, 2
\left\{\, 1- \exp \left[-\frac{1}{2}\,A\,T_A(b)\,\sigma_{dip}^{\mathrm{p}} (\tilde{x}, \,\rr^2)  \right] \, \right\}\,,
\label{sigmanuc}
\end{eqnarray}
where $b$ is the impact parameter of the center  of the dipole
relative to the center of the nucleus and the integrand gives the
total dipole-nucleus cross section for fixed impact parameter. The
nuclear profile function is labeled by $T_A(b)$.  The parameterization for the dipole
cross section takes the eikonal-like form,
$\sigma_{dip}^{\mathrm{p}} (\tilde{x},\,\rr^2)  =  \sigma_0 \,[\,
1- \exp \left(-Q_s^2(\tilde{x})\,\rr^2/4 \right) \,]$, where one
has  used the parameters from saturation model, which include the charm
quark with mass $m_c=1.5$ GeV and the definition $ \tilde{x}=(Q^2
+ 4\,m_Q^2)/W_{\gamma A}^2$. The saturation scale $Q_s^2(x)=
\left(x_0/x \right)^{\lambda}$ GeV$^2$, gives the onset of the
saturation phenomenon to  the process. The  equation above sums up all the multiple  elastic rescattering
diagrams of the $q \overline{q}$ pair and is justified for large
coherence length, where the transverse separation $r$ of partons
in the multiparton Fock state of the photon becomes as good a
conserved quantity as the angular momentum, namely the size of the
pair $r$ becomes eigenvalue of the scattering matrix. 

(iii) \underline{\bf Color Glass Condensate}: The regime of  a CGC is  characterized by the limitation on the maximum phase-space parton density that can be
reached in the hadron/nuclear wavefunction (parton saturation) and
very high values of the QCD field strength $F_{\mu \nu} \approx
1/\alpha_s$ (For a review see, e.g, \cite{VPG}). The large values  of the gluon
distribution at saturation   suggests the use of semi-classical methods, which
allow to describe the small-$x$ gluons inside a fast moving
nucleus by a classical color field. In Refs. \cite{gelis} the heavy quark production  in UPC's has been analyzed in the CGC formalism.  In Ref. \cite{EPJCPER3}, we have improved that analysis using a realistic photon flux and a color field correlator including quantum radiation effects. The input photoproduction cross section reads as \cite{EPJCPER3},
\begin{eqnarray}
\sigma_{\gamma A\rightarrow Q\bar{Q}X}^{\mathrm{CGC}}= \frac{\alpha_{em}e_q^2}{2\pi^2}\int d\rk_{\perp}^2\left[\pi R_A^2\widetilde{C}\,(x,\,\rk_{\perp})\right]
\left[ 1+ \frac{4(\rk_{\perp}^2-m_Q^2)}{\rkn_{\perp}\sqrt{\rk_{\perp}^2+4m_Q^2}}\,{\rm arcth}\,\frac{\rkn_{\perp}}{\sqrt{\rk_{\perp}^2+4m_Q^2}} \right]\!,\nonumber
\label{dsdy_phen}
\end{eqnarray}
where we define the  rapidity $Y\equiv \ln(1/x)=
\ln(2\,\omega\,\gamma_L/4m_Q^2)$ and  $e_q$ is the  quark charge. In Ref. \cite{EPJCPER3}, we obtained the following analytical expression for the color field correlator considering a suitable ansatz for the dipole-nucleus cross section, 
\begin{eqnarray}
\widetilde{C}\,(x,\rk_{\perp})= \frac{4\pi}{Q_{sA}^2(x)}\, \exp \left( -\frac{\rk_{\perp}^2}{Q_{sA}^2(x)} \right)\,,
\label{csat}
\end{eqnarray}
where we have assumed for the nuclear saturation scale, $Q_{s\,A}^2 (x) = A^{1/3} \times Q_{s}^2 (x)$.

Having summarized the theoretical models for heavy quark photoproduction, let us   present the numerical calculation of their total cross section at UPC's. We focus mostly on LHC domain where small values of $x$ would be probed. In the following, one considers
the charm and bottom  masses  $m_c=1.5$ GeV and $m_b=4.5$ GeV ,
respectively. Moreover, for PbPb collisions at LHC, one
has the c.m.s. energy of the ion-ion system
$\sqrt{S_{\mathrm{NN}}}=5500$ GeV. The results are presented in Table \ref{tabhq}. The collinear approach gives
a larger rate, followed by the semihard approach (labels I and II refer to GRV94 and GRV98 gluon pdf's). The saturation model and CGC
formalisms give similar results, including a closer ratio for
charm to  bottom production. Concerning the CGC approach, our
phenomenological educated guess  for the color field correlator
seems to produce quite reliable estimates. Therefore, the photonuclear production of heavy quarks allow
us to constraint already in the current nuclear  accelerators the
QCD dynamics since the main features from photon-nuclei collisions
hold in the UPC reactions.

\begin{table}[t]
\tbl{\it  The photonuclear heavy quark cross sections for UPC's  at LHC. }
%\begin{center}
{\begin{tabular} {||c|c|c|c|c||}
\hline
\hline
$Q\overline{Q}$   & {\bf Collinear} & {\bf SAT-MOD} & {\bf SEMIHARD I (II) } &  {\bf CGC} \\
\hline
 $c\bar{c}$ & 2056 mb & 862 mb &  2079 (1679.3) mb & 633 mb \\
\hline
 $b\bar{b}$ & 20.1 mb  & 10.75 mb & 18 (15.5) mb & 8.9 mb\\
\hline
\hline
\end{tabular}}
%\end{center}
%\caption{\it The photonuclear heavy quark  total cross sections for UPC's  at LHC }
\label{tabhq}
\end{table}

\section{Photonuclear  production of vector mesons} 

Let us consider the
scattering process $\gamma A \rightarrow V A$ in the QCD dipole approach, where $V$ stands for
both light and heavy mesons. The main motivation for using this approach is the easy intuitive interpretation and its successful data description in the proton case. The scattering process can be seen
in the target rest frame as a succession in time of three
factorizable subprocesses: i) the photon fluctuates in a
quark-antiquark pair (the dipole), ii) this color dipole interacts with the
target and, iii) the pair converts into vector meson final state. In the color dipole formalism, the
corresponding imaginary part of the amplitude at zero momentum
transfer reads as \cite{kope,nem}, 
\begin{eqnarray}
{\cal I}m \, {\cal A}_{\gamma A \rightarrow VA}  = \sum_{h, \bar{h}}
\int dz\, d^2\rr \,\Psi^\gamma_{h, \bar{h}}(z,\,\rr,\,Q^2)\,\sigma_{dip}^{\mathrm{target}}(\tilde{x},\rr) \, \Psi^{V*}_{h, \bar{h}}(z,\,\rr) \, ,
\label{sigmatot}
\end{eqnarray}
where $\Psi^{\gamma}_{h, \bar{h}}(z,\,\rr)$ and $\Psi^{V}_{h,
  \bar{h}}(z,\,\rr)$  are the light-cone wavefunctions  of the photon
  and vector meson, respectively. The quark and antiquark helicities are labeled by $h$ and $\bar{h}$
  and reference to the meson and photon helicities are implicitly understood. The variable $\rr$ defines the relative transverse
separation of the pair (dipole) and $z$ $(1-z)$ is the
longitudinal momentum fractions of the quark (antiquark). 

In the dipole formalism, the light-cone
 wavefunctions $\Psi_{h,\bar{h}}(z,\,\rr)$ in the mixed
 representation $(z,\rr)$ can be completely determined using light cone perturbation theory. On the other hand, for vector mesons, the light-cone wavefunctions are not known
in a systematic way and they are thus obtained through models (For a recent detailed discussion see Ref. \cite{Nikolaev}).  Here, we follows the
analytically simple DGKP approach \cite{dgkp:97}, which  assumes
that the dependencies on $\rr$ and $z$ of the wavefunction are
factorised, with a Gaussian dependence on $\rr$. We keep the original parameters of the model. The main shortcoming of this approach is that it breaks the rotational invariance between transverse and longitudinally polarized vector mesons \cite{Nikolaev}. However, as it describes reasonably the HERA data for vector meson production, as pointed out in Ref. \cite{sandapen}, we will use it here.

\begin{figure}[t]
\begin{tabular}{cc}
\epsfig{file=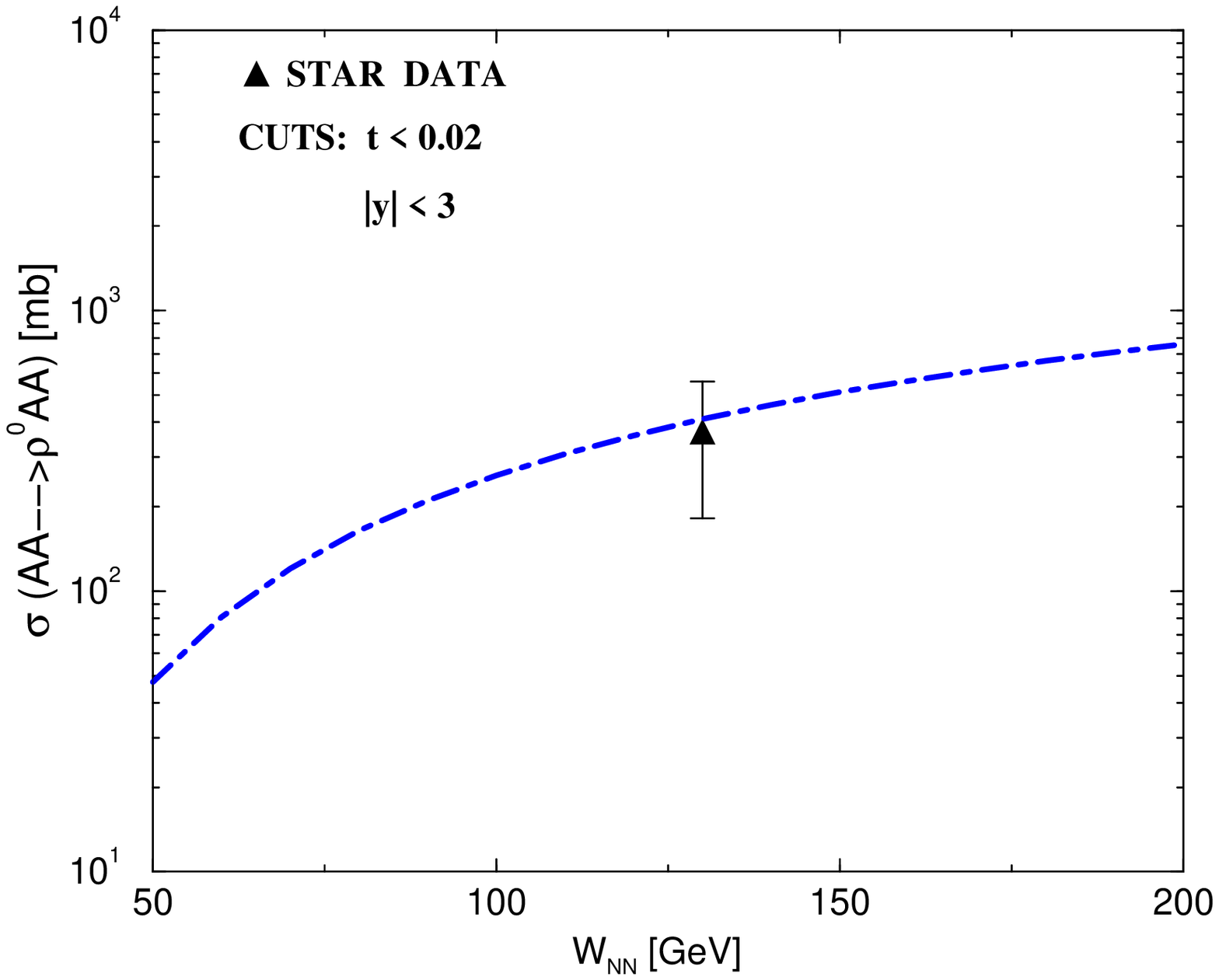,width=62mm} & \epsfig{file=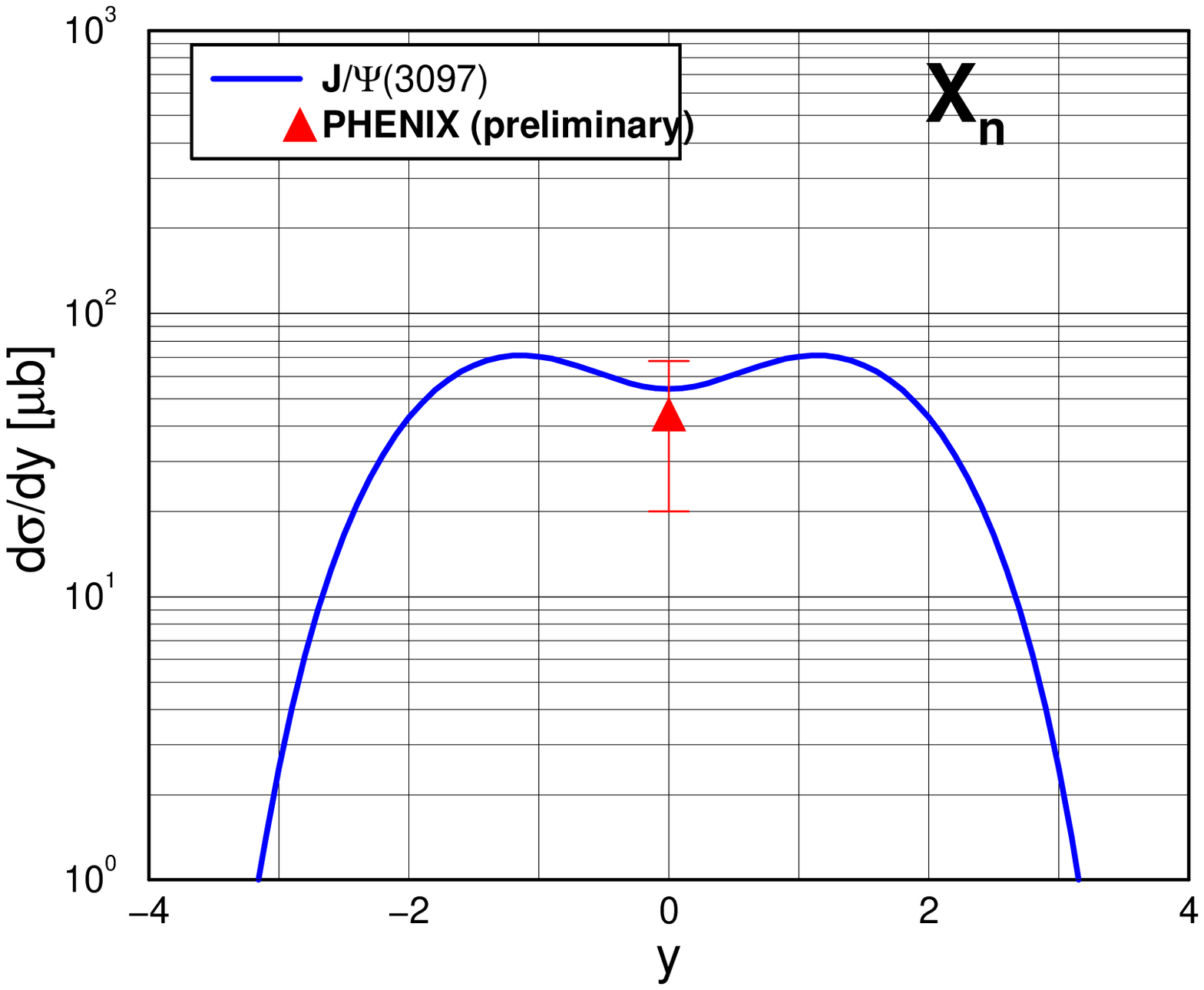,width=62mm}\\
(a) & (b)
\end{tabular}
%\centerline{\psfig{file=star_theor.eps,width=4.7cm}}
%\vspace*{8pt}
\caption{(a) Energy dependence of coherent $\rho$ photoproduction at RHIC. (b) Rapidity distribution of coherent $J/\Psi$ photoproduction (with nuclear breakup) at RHIC (see text). \label{f1}}
\end{figure}

The corresponding parameters for the meson wavefunctions are presented in Table 1 of Ref. \cite{magno_victor_mesons}.  Following Ref. \cite{nem}, 
 we have estimated contribution from real part for the photoproduction of vector mesons: it is about  3\% for light mesons
and it reaches 13\% for $J/\Psi$ \cite{magno_victor_mesons}. Additionally for heavy mesons we have take into account
the skewedness effects, associated to 
off-forward features of the process,  which  are increasingly
important in this case. Finally, the photonuclear cross section is given by
\begin{eqnarray}
\sigma_{\gamma A\rightarrow VA}^{tot} =  \frac{\left[{\cal I}m \, {\cal
      A}_{\gamma A\rightarrow VA}\right]^2}{16\pi}\,\,(1+\beta^{2})\,\int_{t_{min}}^\infty
      dt\, |F(t)|^2 \,,
\label{fotonuclear}
\end{eqnarray}
with $t_{min}=(m_V^2/2\,\omega)^2$. The quantity $\beta$ is the ratio between the imaginary and real part of the amplitude. We have used an
analytical approximation\cite{klein_nis_prc} of the Woods-Saxon distribution as a hard
sphere, with radius $R_A$, convoluted with a Yukawa potential with
range $a=0.7$ fm in order to compute the nuclear form factor, $F(t)$. 

Recently, the STAR Collaboration at RHIC  published  the first experimental measurement of coherent $\rho$ production in gold-gold UPC's at $\sqrt{s} = 130$ GeV \cite{star_data}. The energy dependence of the cross section is presented in Fig \ref{f1}-a. Our theoretical prediction in the curve take into account the experimental cuts, which gives $\sigma_{\mathrm{theory}}(|y| \leq 3)= 410$ mb, in good agreement with the STAR measurement $\sigma_{\mathrm{STAR}}(|y| \leq 3)= 370\pm 170\,(\mathrm{stat}) \pm 80 \,(\mathrm{syst})$ mb. Furthermore, the PHENIX Collaboration has preliminary measurements of the cross section for coherent $J/\Psi$ photoproduction in UPC at midrapidity at $\sqrt{s} = 200$ GeV \cite{phenix_data}, accompanied by nuclear breakup. The rapidity distribution is shown in Fig \ref{f1}-b in comparison with our prediction. The theoretical estimation gives $d\sigma/dy|_{y=0}= 58$ $\mu$b, which is in agreement with the PHENIX  measurement $d\sigma_{\mathrm{PHENIX}}/dy|_{y=0}= 48\pm 16 \,(\mathrm{stat}) \pm 18 \,(\mathrm{syst})$ $\mu$b. Finally, in  Table \ref{tab1} one shows the predictions for the integrated cross sections for LHC. The experimental feasibility and signal separation on the reaction channels presented here are reasonably clear, namely applying a (low) transverse momentum cut $p_T<1$ and two rapidity gaps in the final state for the meson case. In contrast, for heavy quarks we have only one rapidity gap.

\begin{table}[t]
\tbl{\it Cross sections for coherent meson photoproduction at LHC energy. }
%\begin{center}
{\begin{tabular} {||c|c|c|c|c|c||}
\hline
\hline
& {\bf HEAVY ION}   & $J/\Psi\,(3097)$ & $\phi\,(1019)$ & $\omega\,(782)$ & $\rho\,(770)$  \\
\hline
\hline
 {\bf LHC} & CaCa &  436 $\mu$b & 12 mb & 14 mb &  128 mb \\
\hline
&  PbPb &  41.5 mb &  998 mb & 1131 mb & 10069 mb \\
\hline
\hline
\end{tabular}}
%\end{center}
%\caption{\it Cross sections for nuclear vector mesons photoproduction at UPC's at RHIC and LHC energies.}
\label{tab1}
\end{table}

\end{document}